# Modeling a repository of modules for ports Terminals Operating System (TOS)

Ahmed Faouzi[1], Charif Mabrouki[2], Alami Semma[3]

[1] Department of Mathematics and Computer Science, Faculty of Science and Techniques,
Hassan 1 University
Settat, Morocco
*ahmed.faouzi@gmail.com*

[2] Department of Mathematics and Computer Science, Faculty of Science and Techniques,
Hassan 1 University
Settat, Morocco
*charif.mabrouki@cigma.org*

[3] Department of Mathematics and Computer Science, Faculty of Science and Techniques,
Hassan 1 University
Settat, Morocco
*semmaalam@yahoo.fr*

**Abstract**
The purpose of this paper is the modeling of a repository for modules and interfaces that must include all integrated information system management of a port terminal.

Modules will provide a basic framework necessary for automatic management of internal operations and activities of all Port Terminals worldwide.

Interfaces will provide a basic framework for the management of external operations subject to standardized exchange with partner's Information Systems of the port community.
These modules and interfaces will be used for the implementation of all integrated systems for the management of a Terminal Port (TOS).

## 1. Introduction

In all countries the Port is located as a core around which we find a community in general called "Community Port": Port Operators, Port Authority, Customs, Bank, Officer Maritime Carrier, Clients ...
Overall, we can divide the activity of a port into two categories: Container traffic activity (goods carried in containers) and various traffic activity (goods made in other types of packaging: cereals, oil, wood ... ).

Information System plays a very important role in the activity of a Terminal Port, firstly to ensure business operational in Port (Goods and Treatments Ships) and secondly to ensure the interface with members of the port community.
In Market of Port's Information Systems there is a lack of a common repository of all modules that must implement integrated information system for the management of container traffic, Also there is a lack of integrated system for the management of Various traffic.
This article addresses the modeling of a common repository of modules and interfaces that must include all information system for managing container traffic.

## 2. Problematic

✓ Lack of a common repository of modules that must include all integrated information system for Container Port Traffic,

✓ Difficulty to have a repository for electronic exchange interface to community's Information Systems

## 3. Terminal Port Activities

### 3.1 General Services

Below the mains activities in a Port Terminal :

✓ Services to goods handling aboard ship and shore handling : Container, Various good,

✓ Services to Ships: Pilotage, Towing, providing water / energy ...

✓ Storage, scoring, weighing, packing and unpacking of containers and trailers

✓ Other Services: hauling, stacking of goods, loading and unloading trucks;

### 3.2 Operating Procedures (Container Traffic)

#### 3.2.1 Import process:
1. Arrival of the ship at the dock
2. Unloading import containers
3. Storage terminal
4. Delivery of containers

For this process we will be considered the following cycles:
• Cycle truck
• Cycle ship
• Movements in full land

#### 3.2.2 Export process:
1. Arrival of the truck delivering the export container
2. Storage terminal
3. Loading the container on the vessel.

For this process we will be consider the following cycles:
• Cycle truck
• Cycle ship
• Movements in full land

#### 3.2.3 Trucks Cycles (Import, Export)

Before the physical input of the truck terminal, paperwork must be completed, and the order "service request" delivery container inquired about the system, customs clearance must be paid.

Orders and authorizations can happen in paper format or in EDI messages.

Before the physical input of the truck, port terminal in general practice administrative control, and instruction document is generated for the truck, then follow the physical check.

Physical Control definition:
Physical control is defined as the visual inspection by a pointer that verifies container numbers, license plate information CSC (Container Safety Certificate), stickers IMO, presence and seal numbers, or any other inspection that may be required by the operator or by the contract between the operator and shipping line or other customer.

If the physical check is passed, the truck is directed to the interchange area designated by the business Information System.

In the interchange, the container will be unloaded or loaded on the truck. Of course, a trailer can load multiple containers, and therefore in the interchange area we can handle multiple containers.

Movements combined I / O must also be possible.

After treatment of the truck, it proceeds to the exit door.

#### 3.2.4 Ship Cycle: Import

1. Message BAPLIE: the terminal receives EDI BAPLIE plan. This BAPLIE will be associated with the ship's visit. Based on a list of containers BAPLIE discharging that may be generated.
2. Arrival of the ship after ship registration in the Information System and planning the sequence containers and cranes, unloading the ship can begin.
3. After execution of movements, the container reaches its planned position. The movements are recorded using a laptop or RTD or VMT.
4. After handling all containers the ship's visit will be closed. Handling means : all the operations by ships (import and export).

#### 3.2.5 Ship Cycle: Export

1. Message MOVINS: the terminal receives EDI MOVINS plan. This plan MOVINS be associated with the ship's visit. Based on the presence and MOVINS containers in the terminal a list of containers to be loaded can be generated. MOVINS message contains instructions indicating the owner of the bays in the ship for some types of containers.
2. Arrival of the ship after the registration of the ship in Information System and planning the sequence containers and cranes, vessel loading can begin.
3. After execution of movements, containers reach their planned position in the ship. Movements are recorded by means of a portable computer,
4. After handling all the export and import containers, the ship's visit will be closed.

**4 Modeling a repository of the Basic Modules required for an Integrated Information System of Port Terminal (TOS : Terminal Operating System)**

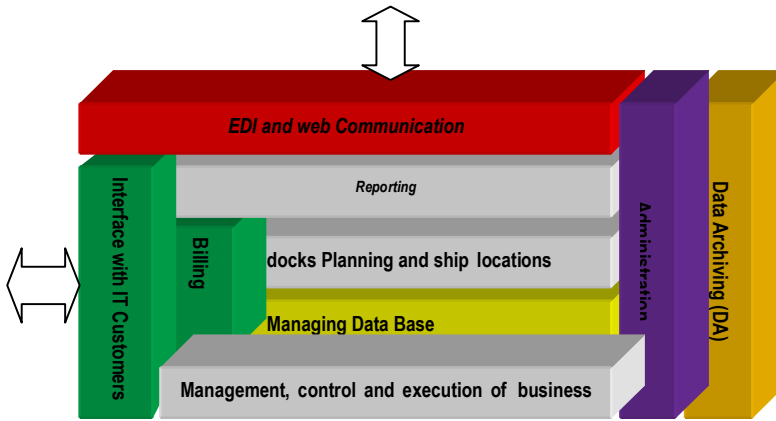

**Modeling TOS System modules**

4.1 Module 1 : docks Planning and ship locations:
The TOS (Terminal Operating System) must integrate these modules:
- Planning Docks (Location vessel)
- Allocation of Human resources and equipments resources (the ship)
- Planning Ships
- Planning full land
- Planning and Control gear

4.1.1 Planning docks and Resource allocation
The planning docks module and ship locations (BP, Berth Planning) allow to plan (at short and at long term) the arrival and work of the ships "human and equipments resources'. It must give a correct view in real time about the current ships arrival and operative portal vessels.
- BP module consists of an array of planning. One axis represent ships arrival and second axis represents time. The BP should allow seeing the current time, the history and future situations.
- The physical data of each ships arrival must be registered in the BP system: water depth, maximum vessel tides;
- BP module simulates the effect of attracting additional customers, and whether the terminal has capacity to dock, where this ability is? and when this capacity is available ?

General objective: the planning should be in order of priority:

1. The gantry will work to maximum rate during loading or unloading. This means delivery / Evacuation containers synchronized with the rhythm of the gantry.
2. Distances traveled by vehicles should be as short as possible
3. Port Machines should not circulate without container: no unproductive trips
4. The capacity utilization of machines, including RTG machines should be balanced. This means that the number of movements assigned to RTGs will be more or less equal by RTG.
5. The number of unproductive movements should be as small as possible: the TOS must avoid "Shifting".
The basis for effective operations is the availability of accurate information before and during the operational cycles.
Computer systems must have interfaces in order to receive information from business partners and authorities. We mean by the authorities port authority, harbor and customs.
All movements of containers on the terminal are programmed and planned on the basis of information on container handling: orders for loading, unloading, transfer visit, delivery or receipt portal ....
All planning must be based on the "next container move", to avoid non-productive movements in the terminal. For example, planning a container transshipment received is based on trip data with which the container leaves the terminal.
The TOS should plan the container so that the inverse image of the planning will be obtained on the ship terminal. This prevents repositioning unproductive.

The TOS should allow the use of various planning strategies on full land, depending on the composition of stacked containers and means of transport
The general principle to plan the locations of containers is as follows:
1. The operator defines filters to group containers in collections. The criteria to filter the containers can be: container size, weight, service, port of loading, port of destination...
2. The operator sets for containers belonging to a collection where they will be stored, so the operator defines storage areas. The definition must be able to perform each type of transaction (import, export, transshipment), by service ship or a ship visit.
3. Defined area or by the whole terminal, the operator defined how to stack containers by area: containers heaviest over lighter containers, containers for the same destination port on the same destination port, containers for the same container ship on the same ship ...

The operator must therefore assign a storage area for each collection. In the case where a container cannot be "filtered" in a collection, the operator must manually plan the container

Regarding the strategy to assign zones collections, editor must implement at least the following strategies (based on logical storage areas in the TOS or virtual, not on physical breakdown on the terminal):

Strategy 1: Group all containers belonging to the same collection in an area.
This option makes sense and is a good strategy for Import Operations. This strategy puts a lot of pressure on the RTG machine, it must handle a large number of containers, which will limit the performance of the gantry as the technical performance of an RTG is less than the technical performance of a gantry, the RTG will not keep pace with the gantry. In addition this strategy has a consequence: in case of failure of the RTG machine all operations may stop.
Strategy 2: Distribute containers belonging to the same collection to areas distributed: Scattered Planning (SC)
In the case of SC, the system distributes the containers through several zones. Consequently, the amount of work or number of movements will be divided by number of RTG machine. This is logical, because the technical performance of an RTG is lower than the performance of a gantry. The TOS must ensure that the distances between gantry and RTGs is not too high because it will meet the second principle of the Strategy 1.
Strategy 1 is considered the best for a terminal in which traffic Import premium
Strategy 2 is considered the best for a transshipment terminal.

Planning and storage IMO containers (International Maritime Organization) on container terminals is based on the rules of segregation IMO and if applicable, on the national or local rules.
IMO (International Maritime Organization) rules classify containers, based on their content in classes. For each class, rules of segregation were defined by the IMO.
These rules essentially segregation specify if a minimum distance must be maintained between the containers belonging to certain categories of IMO, if the containers can be stored on the terminal, etc.. These containers will be routed to the dedicated slots for storage containers IMO.
Containers "non-waterproof" are another special case. For these containers dedicated physical installation is required.

Operational procedures associated with planning strategies allow the terminal benefit from the expertise of TOS.

4.1.2 Planning Ships (PS)
The TOS must allow planning ships as follow:
- Registration of ship characteristics: physical dimensions, capacity, crane positions, cockpit, stability data and requirements in order to optimize the loading of the ship, heights and maximum weight per container stack.
- Integration of EDI messages (BAPLIE and MOVINS) editing and corrections if necessary EDI messages before applying them on the ship's structure to processing
- Simulation of different work scenarios. The operator will receive information on the effectiveness of the calculated planning systems. The operator can adjust the settings accordingly by planning, taking into account the parameters and stability requirements. Final planning will be stored on CD or USB stick and will be transmitted to the master for verification and approval
- Assignment of an optimal number of gantries to the ship in order to achieve the contractual terms of the handling operation to finish before the ETD (Estimate Time Departure) of the ship.

PS module will consider the positions of the containers terminal and minimize repositioning of the terminal and reduce the distance by port equipment.
PS module will consider in planning the availability of gantries and port equipment and notify operators of resources to achieve handling in a time.
The module should include the ability to plan and manually change the schedule for automatic containers selected by the operator.
The module will consider the IMO rules, adapted by operators during the planning of ships.
PS module allows the monitoring in real time all operations vessel.

4.1.3 Planning full land (PL):

The TOS should allow the distribution of terminal storage areas following names adapted to the requirements of the operator: buildings, roads, intersections, arrival ship, areas Reefers portals, Interchange area, and any other structure on the terminal.

The operator must have functions to modify the plan of the terminal in real time and without the intervention of the software. PL module allows you to define areas not available due to a scheduled maintenance.

PL module should allow the definition of rules for the planning of containers on the terminal based on different criteria such as:

• Fixed Fields import, export
• Hotspots next service "vessel call" etc…
• Way storage container
• Strategies for group containers following example weight, port of destination, consignee
• Storage Strategies to avoid port congestion gear
• Storage Strategies to reduce distances by the craft port.

PL module should have functions for:
• Show real-time capacity on the terminal
• The future capacity of the terminal
• Adapt the rules for selecting storage areas, as well as how to stack containers in real time and to notify the operator before unloading if, based on simulation, the rules are not sufficient to storing the containers.
• The recording positions of containers using GPS or similar
PL module provides functionality "drag and drop" to manually change the positions of containers.
PL module provides functionality to reorganize the terminal operations to prepare the ship loading or unloading and trucks operations.

PL module communicates in real time with other TOS modules, and contains modules for specific Reporting planning positions of containers. The module will consider the IMO rules adapted by operators.
5.1.4 Planning and control gear (PG):
Module planning and control gear allows the detailed specification of all types of port equipment such as RTG machines, the reach stacker, trucks, trailers.. with all their technical characteristics and handling capacity (maximum stacking height, maximum speed vacuum speed with full container, lifting capacity, GPS sensors and weighing..).

The definition of gear also contains data on the patterns movement and links between machines in the circuit control PG module connections between individual movements and synchronizes the steps most effectively "Example To move a container, in the first RTG machine must load the container on a chassis, The chassis must be transported by a tractor to the gantry, The gantry will unload the container from the chassis"

The PG module transmits instructions to terminals RTD Radio Frequency (Radio Data Terminal) installed on the gantry. The driver will confirm the execution of an instruction and receive the next instruction.

The PG module contains the history of all the movements of each device port, the duration of each movement as well as dependencies.

PG module evaluates in real time the position of movements and optimizes the individual instructions to be made and the use of all equipments.
In case the machine wills no instructions, the system informs the operator immediately so that they can assign equipments to work when needed. PG module allows to assign equipments to specific operations.

The workflows are automatically assigned according to deadlines or manually by operators. The PG system receives other sequences from the planning vessel module.

The PG module allows generating internal reports productivity and provides an interface with CMMS (Computer Maintenance Management Systems).

4.2 Module: Managing Data Base (Kernels)
Data bases (DB) should be centrally managed and controlled, and basic data will be used by all modules. Data such as a directory of ports, ISO containers codes …
The TOS must include categories of data:
1. Commercials partners: ships owners (operators of ships, NVOCC Non-Vessel Operating Container Carriers), shipping lines, agents, trucking companies; any other party may receive invoices for port terminal services.
2. Transport Data: ships, fields, trucks, Ship structure, such as physical characteristics, weight…
3. Maritime services and port rotation
4. Ports directory following conventions
5. Container directory types according to ISO rules
6. Terminal Operators
7. Work shifts, hours of operation standard and overtime
8. Definition equipment and technical
9. Staff
10. System users: roles, access authority (editing, creation, consultation) and functional safety
11. Overall system configuration
12. Terminal Definition

4.3 Module: Data Archiving (DA)
This DA module should include Archiving Strategy:
The purpose of archiving is to maximize the system's disk space of the TOS and also optimize TOS efficiency and response time.
The archival frequency times depend on legislation in the country.
The DA module must contain 3 archiving strategies:
1. Operational archiving (data date: 2 to 3 years): Data are available on the operating system and are available through the user TOS interface.

2. Mid-term archiving (data Date: 3 to 10 years): Data are available on a medium that is not part of the operating system, but can be consulted by the user TOS interface.
3. Long-term archiving (date of data > 10 years): Archiving is done in an outside medium searchable by SQL or TXT tools.

5.4 Module: Management, control and execution of business processes
This module must include:
✓ Administration of orders
✓ Monitoring and execution

The execution part is an integral part of operational modules as described in the previous pages. This module manages a comprehensive operations and procedures of container terminals Mainly it provides all the user interfaces of recording, and allows the operating procedures. All cycles, links with the authority, customers and partners are scheduled at this level.

4.4.1 Administration orders (AD)
Administration orders <or Services request > allows to create, modify and cancel orders if necessary.
Orders are in effect requests services from partners. Service requests are typically the list of services to clients:
1. Receiving a full container delivered by truck
2. Delivery of an empty container to a truck
3. Full container delivery to a truck
4. Receiving an empty container
5. Request to load a container on a ship
6. Request of unloading a container from a ship
7. Request to receive a ship terminal, arrival time, departure time,...
8. Request for verification of lead
9. Demand for warehousing of lead line shipping
10. Application for the condition of a container
11. Application control and warehousing stickers IMO
12. Request CFS operations
13. Request fumigation, cleaning etc..
14. Demand for transportation equipment repositioning or land full
15. Change request travel / service vessel
16. Inspection request (initiated by the customs)
17. Asked to provide specific services to vessels
18. Preparation request refrigerated containers (Pre-tripping)
19. For reorganization cargo ship (Restows)
20. The ability to register and manage bottlenecks imposed by customs or other authority, interface with Customs

In principle, any service provided by the port terminal must be registered in the system TOS. Registration application is the basis for the planning, implementation and ongoing operations.

4.4.2 Monitoring and Execution: (ME)
All automation process and management for these orders must be implemented in the TOS, in this part will be integrated orchestration, editing interchange document, compliance monitoring, management rules…

The ME module is integrated and interfaced with planning module, EDI communication, WEB Client Access, Reporting, and billing.

The module must allow the execution of any instruction generated by the TOS in real time and plan subsequent movements based on the confirmation of execution of the previous movement.

Examples:
1 / for the AD order on the Reefers This ME module will provide interfaces to allow RTD terminal to receive instructions for connecting and disconnecting refrigerated. The interfaces must also input temperatures of refrigerated; it will return data to the client.
2 / For the vessel treatment, ME module will integrate automatically EDI messages like : Bayplan MOVINS, PRESTOW. It will provide user interfaces for stops input, board and land score, orchestration modules with planning, communicate orders for tractors and RTG, validate positions containers, refunds performance handling, editing documents and reports, return EDI messages like : COARRI, CODECO (if direct outputs), …

4.5 Module: Communication and web access
✓ EDI (Electronic Data Interchange) Communications
✓ Interface WEB Extranet

The TOS must integrates EDI communication modules (Electronic Data Interchange) and WEB.
The EDI module should allow the exchange of structured messages between the port terminal and its partners and can handle all standard EDIFACT messages used in practice in the field of handling and processing of container terminal (Import / Export / Transshipment).

The EDI module should define the partners with which the port terminal communicates. The EDI module allows you to modify EDI messages according to need and standards in accordance with the partners of the port terminal.

The EDI module must have a web interface allowing partners to not only view data on their operations, but the interface must also allow partners to record data and orders (service requests).

WEB features contain all devices to ensure a good systems protection: applications, databases, users profiles, access permissions to some or several TOS modules and data

4.6 Module: Reporting (RE)
4.6.1 Reporting of each module
Each TOS module must have its own Reporting functions. The TOS should have standard reports and must also allow authorized users to create their own reports.
The TOS should allow the printing of reports and the saving reports in standard formats (Microsoft Office ®) and sending reports by e-mail.
Reports should be able to compile information from various TOS modules.
As Port Terminal must have the ability to generate their own reports using a reporting tool or Business Intelligence (BI) of his choice, the TOS must provide complete documentation of the TOS database.
for a real-time Reporting, the TOS must have modules showing in real time the status of operations. This module allows users to define their own key performance indicators (KPI).
This module must generate views of the terminal indicating the performance by area and by operation type. The user can click a view details for domain, area of operation, equipment movement.
RE module must be available through an internet connection (HTTPS) for only authorized users. The terminal can manage users without the intervention of the TOS editor.
Port Authority (under the agreement) requested statistics like :
- ✓ Volumes handled,
- ✓ Dangerous Materials,
- ✓ Work accidents,
- ✓ Productivity and performance according to the guidelines and definitions established by Port Authority.

4.7    Module: Billing (BI)

The system must allow TOS to charge all customers services according to the contracts and prices recorded in the TOS.

- ✓ Billing by customer contract
- ✓ Billing Services recorded in the TOS
- ✓ Billing Services not registered in the TOS
- ✓ Sending Electronic Invoice
- ✓ Generate Report and Statistics

Billing module allows recording all customer data and all identifiers according to the country concerned.
Contracts must be recorded in the module BI and contain at least:
- Types of services to be billed,
- Validity of the contract, discounts, ...
- Benefits
- Rates
- Periods
- Discounts

The BI module allows managing a standard contract (default) which will be used if there is no contract for a client or a group of clients.

BI module allows billing in advance and receive advances. The BI will consider the advances at the time of the creation of the final invoice.
BI modules allow billing occasional services which are not recorded in the operational modules of the TOS. BI modules can send invoices in an electronic format or by EDI.
Functions statistics and reports are available in the BI modules.
The BI module allows analyzes based on cost and revenue referring to customers contracts, and allows evaluating the profitability of currents and futures customers.

4.8 Module: Interfaces with Information System of Customer
In general several TOS clients have already their own billing Information System and they will not use the integrated BI TOS module so TOS should allow a billing interface with this Information System,
Exchanges between the Billing Information System of Port Authority and the Information System of terminal (TOS) is mainly related to billing requirements and integration management control, the TOS will provide to the Billing Information System all data needed to billing :
- ✓ Handling
- ✓ Storage
- ✓ Special Operations
- ✓ Vessels and containers services

The system shall provide to the Billing Information System an access to customer data, specific contracts, to allow customization of billing.
The system will provide in real-time a file data interface, and will provide to the billing information system special orders to unlocks/locks "customers, container"
The system will manage the invoice sequence number between the Billing Information System and the TOS module BI.

4.9 Module: Administration

This module must include:
- Create Users,
- Create Groups of users
- Create Roles with Modules and sub modules
- Access Control modules
- Control access to the database (see Fields)
- Logging Operations
- Module Performance and High Availability
- Load Sharing Mode
- Backup Mode
- Error Handling
- Administration and Supervision
- On line documentation

## 5    Conclusion

This work was performed on findings observed in the field of the Port Terminal and following loan business operations at the Port.

We met several difficulties along our realization, e.g.: access to business data, process development activity, the release of critical activities, research common modules between the ports of the world...

The common repository described in this article, which set the modules that must incorporate all Port Information System is the result of our work, firstly this work gives a hand to any editor -wishing to invest in port's Information Systems development- by having a repository of basic modules to implement, and secondly this work gives to customers 'Port Operators' buyers of integrated software solution a repository for help choosing of a market solution.

Modeling modules for an information system of various traffic is a perspective that can be the subject of another research.